\documentstyle[titlepage,fleqn,11pt]{article}
\begin{document}
\begin{titlepage}
\title{M\"{o}bius Inverse Problem for Distorted Black Holes }
\author{H.C. Rosu\\
Instituto de F\'{\i}sica de la Universidad de Guanajuato\\
Apdo. Postal E-143, Le\'on, Guanajuato, M\'exico\\
}
\date{November 1995}
{\baselineskip=20pt
\begin{center}
{\bf Abstract}
\end{center}

Hawking ``thermal'' radiation could be a means to detect black holes
of micron sizes, which may be hovering through the universe. We consider
these micro- black holes to be distorted by the presence of some
distribution of matter representing a convolution factor for their
Hawking radiation. One may hope to determine from their
Hawking signals the temperature distribution
of their material shells by the inverse black body problem.
In 1990, Nan-xian Chen has used a so-called modified M\"{o}bius
transform to solve the inverse black body
problem. We discuss and apply this technique to Hawking radiation of
distorted micro-black holes.
Some comments on supersymmetric applications of M\"{o}bius function
and transform are also added.

\vskip 2cm
PACS numbers: 04.60 , 04.20.

MCS numbers: 83C57.

\vskip 1cm
Corrected and updated draft of
Nuovo Cimento B 108, 1333-1339 (1993)

\vskip 1cm
gr-qc/9511013

    }

\vskip 1cm

\maketitle
\end{titlepage}
\baselineskip=30pt
\parskip=0pt

\section{Introduction}

Recently, the remarkable 160 years old inversion/transform formula of
M\"{o}bius has vigorously entered Physics through a number of
important applications to inverse problems \cite{03}. Its future in
mathematical physics and even in more applicative science seems to be
extremely fruitful \cite{01},
although Hughes, Frankel and Ninham \cite{02} have shown
that M\"{o}bius function occurs only because the expansion of the
reciprocal $\zeta$ function in Mellin transforms produces it.
 The interest in M\"{o}bius inversion was raised by
Nan-xian Chen, who proved a modified M\"{o}bius transform in
order to apply it to such problems like finding out
the phonon density of states, the inverse
black body radiation problem, and getting the solution for inverse
Ewald summation. In an astrophysical context the M\"obius transform
has been applied to the analysis of {\em IRAS} data of interstellar
dust emission and also to the star-forming condensations in the core of
several dense molecular clouds \cite{Xie91}.

The present work is prompted by the inverse black body problem
in the realm of black hole physics. Hawking heuristic discovery of
the black body radiation from Schwarzschild horizons is a famous result
of quantum field theory in curved space-times \cite{2}.
Consequently, one might be well motivated to look for primordial black holes,
and more generally,
for small (mini) black holes, from the point of view of remote sensing,
i.e., the determination of the ``surface'' temperature distributions
from the spectral measurement of their Hawking radiation. We have in mind the
so-called {\em distorted black holes} discussed by Geroch and Hartle
\cite{GH82}. These are Einstein vacuum solutions obtained by the Weyl
technique
of generating solutions from axially symmetric potentials of the flat
space Laplace equation. Weyl black holes are stationary and
carry with them an external distribution of matter; they can have only
spherical or toroidal horizons. In this
astrophysically interesting class of objects, one may include
mini-black holes
surrounded by thin matter shells \cite{Br91}. We call the {\em horizon
temperature distribution} the quantity conjugated to
the Hawking power spectrum through the inverse transform but this
obviously does not mean we assign the distribution directly to the
horizon surface. In the case of {\em distorted black holes} because of the
matter distribution the effective horizon temperature is not constant
and therefore it is meaningful to think of a sort of surface temperature
distribution. There is no restriction imposed by the zeroth
law of black hole thermodynamics \cite{bch73} which applies only to
{\em isolated} static or stationary black holes.
I recall that the generalized
black-hole thermodynamics has been settled 20 years ago as a
consequence of Hawking radiance, but this in turn was the starting
point for still greater puzzles.

%The interior (better surface/horizon)
%microstates
%are quantum correlated with the outside fields in the right way to
%produce black body radiation in each quantum mode.

\section{Inverse problem for Planck law}

In what follows we shall consider the Planck radiation in the
radiometric- remote sensing approach.

What is known as Planck law is the analytical formula for the
power spectrum or spectral brightness of the black body radiation
\begin{equation}\label{eq:b1}
P(\nu,T)=\frac{2h\nu ^{3}}{c^{2}}
\frac{1}{\exp (\frac{h\nu}{kT})-1}\;.
\end{equation}
In radiometry the power spectrum is called spectral radiance, and
characterizes the source spectral properties as a function of position
and direction. The total radiated power spectrum (which in radiometry
is called radiant spectral intensity to be used for point, i.e.,
far away sources) reads
\begin{equation}\label{eq:b2}
W(\nu)= Const\cdot \int_{0}^{\infty} A(T)B(\nu,T) dT\;\;,
\end{equation}
where the area-temperature distribution is denoted by $A(T)$, and the
Boltzmann-Planck factor by $B(\nu,T)$. The inverse black body problem is to
solve the integral equation for $A(T)$ for given total radiated power spectrum,
which may be known experimentally or otherwise. This problem was solved
by Bojarski \cite{b}, who introduced a thermodynamic coldness
 parameter $u=h/kT$, and
an area coldness distribution $a(u)$, as more convenient variables to get
an inverse
Laplace transform of the total radiated power. The coldness distribution
is obtained as an expansion in this Laplace transform.
Later, people working in
the field have given a simpler approach and tried to improve in various
ways the result of Bojarski.

More precisely, the total power spectrum  is rewritten as
\begin{equation}\label{eq:b3}
$$W(\nu)=\frac{2h\nu ^{3}}{c^{2}}\int_{0}^{\infty}
\frac{a(u)}{\exp(u\nu)-1}du\;.
\end{equation}
A series expansion of the denominator leads to
\begin{equation}\label{eq:b4}
W(\nu)=\frac{2h\nu ^{3}}{c^{2}}\int_{0}^{\infty}  \exp (-u\nu)\cdot
\sum_{n=1}^{\infty} (1/n)\;a(u/n)du\;,
\end{equation}
and we already see that the sum
\begin{equation}\label{eq:b5}
f(u)=\sum_{n=1}^{\infty}(1/n)\;a(u/n)\;,
\end{equation}
is the Laplace transform of
\begin{equation}\label{eq:b6}
g(\nu)=\frac{c^{2}}{2h\nu ^{3}}W(\nu)\;.
\end{equation}
{}From the Laplace transform f(u), Chen obtained a(u) by his modified
M\"{o}bius expansion,
\begin{equation}\label{eq:b7}
a(u)=\sum_{n=1}^{\infty} \frac{\mu(n)}{n}f(u/n)\;,
\end{equation}
reproducing in a simple way a result already obtained by Kim and
Jaggard \cite{kj}.

\section{M\"{o}bius expansions}

The classical M\"{o}bius expansion refers
to special sums (which one may call d-sums) of any number-theoretic
function $f(n)$, running over all the factors of $n$, 1 and $n$ included. Such
a kind of running is indicated by the symbol $d|n$.
If
\begin{equation}\label{eq:b8}
F(n)=\sum_{d|n}^{n} f(d)\;,
\end{equation}
then the last term of the sum, $f(n)$, could be written as a sum of
$F$ functions, which we shall call M\"{o}bius d-sum
\begin{equation}\label{eq:b9}
f(n)=\sum_{d|n}^{n} \mu(d) F(n/d)\;,
\end{equation}
in which $F(n)$, i.e., the previous sum, becomes the first term.
Since each $F$-term in the second expansion is a d-sum, it is clear that
we have an overcounting unless the factors $\mu(d)$ (M\"{o}bius
 functions)
are sometimes naught and even negative. The partition of the prime factors
of $n$ implied by the M\"{o}bius function is such that $\mu(1)$ is 1,
$\mu(n)$ is $(-1)^{r}$ if $n$ includes $r$ distinct prime factors, and
$\mu (n)$ is
naught in all the other cases. In particular, all the squares have no
contribution to the M\"{o}bius d-sums, and all the prime numbers have
negative contribution. Another terminology is to call squarefree the
integers selected by means of the M\"{o}bius function.

Chen's trick is to apply such a multiplicative inversion of finite
number-sums
to continuous functions expressed as particular infinite summations
(see below).
If
\begin{equation}\label{eq:b10}
f_{1}(x)=\sum_{n=1}^{\infty}f_{2}(x/n)\;,
\end{equation}
then,
\begin{equation}\label{eq:b11}
f_{2}(x)=\sum_{n=1}^{\infty} \mu(n)f_{1}(x/n)\;.
\end{equation}
This is the modified M\"{o}bius transform (MMT) of Chen. For the
inverse blackbody radiation, $f_{1}(x)=uf(u)$ and $f_{2}=ua(u)$.
So one gets the coldness distribution simply by multiplying the Laplace
transform of the total power spectrum by the coldness parameter and then
applying the MMT (Chen's trick).

\section{Distorted black holes and their coldness distribution}

We come now to some well-known black hole results. In his note
to {\em Nature} in 1974,
Hawking argued that for Schwarzschild black holes (SBHs) the number of
particles emitted in wave packet modes propagating along null u-lines is
$[\exp(2\pi \omega /\kappa) -1]^{-1}$ times the number of particles that
would have been absorbed from similar wave packets incident on the
black hole from $I^{-}$ along null v-lines,
 a result that holds for a perfect black body
 with a thermodynamic temperature in geometric units of $\kappa/2\pi$.
 According to Hawking, the rate of emission for a given bosonic mode is
 $\Gamma _{n} B(\nu, T_{H})$, where $\Gamma _{n}$ is the absorption
 coefficient for the mode, and $B(\nu , T_{H})$ is the Boltzmann-Planck
 factor. Calculations of absorption coefficients were provided by
 Page \cite{p},
 and in the case of massless scalar fields by Matzner \cite{ma68} and
 more recently by Sanchez \cite{s}.
  She was able to
 show that s-wave contribution predominates in Hawking radiation of the
 massless scalar fields. The contributions from all the higher partial
 waves are supressed by the rapid decrease of the Boltzmann-Planck
 factor in the range $kr_{S}\geq 1$. At the same time a strong increase
 of the maxima of the effective potential was found for partial waves
 with $l\geq 1$. The Hawking emission regime in the range
 $kr_{S}\leq 1$ corresponds to
 only phase shifts in the phase of the wave packets passing through the
 Schwarzschild sphere.

Consider now micron-sized SBHs ($M\sim 10^{24}$ g) for which no known
massive particles are emitted, and suppose it radiates in the Hawking
regime. According to Page, about $16\%$ of its Hawking flux goes into
the photon flux, the rest being neutrino emission. Suppose now that
such SBHs are of Weyl type. From the remote sensing point of view we
may introduce
a horizon coldness parameter $u_{S}=h/kT_{DH}=4\pi/\kappa _{D}$,
where $T_{DH}$
is the effective horizon temperature of the distorted black holes, i.e.,
\begin{equation}\label{eq:b12}
T_{DH}=(8\pi M)^{-1}e^{2{\cal U}}\;.
\end{equation}
The variable ${\cal U}$ is related to the exponents that characterize
the Weyl metrics, which one can put in a general form as follows
\begin{equation}\label{eq:h1}
g^{ab}=e^{2U-2V}h^{ab}+r^{-2}e^{2U}(1-e^{-2V})\phi ^{a}\phi ^{b}-
e^{-2U}t^{a}t^{b}\;,
\end{equation}
where $h^{ab}$ is a flat, positive-definite, three-dimensional metric,
$\phi ^{a}$ is the rotational Killing field of the metric, and $t^{a}$
are surface-orthogonal Killing fields. With this notation, ${\cal U}$
will be
\begin{equation}\label{eq:h2}
{\cal U}=U-\frac{1}{2}V-\ln(\frac{1}{2}\sqrt{r/M})\;.
\end{equation}
In the case of Weyl distorted-black holes a simple
application of the MMT will give for the horizon coldness distribution
the expression
\begin{equation}\label{eq:b13}
a(4\pi/\kappa _{D})=\frac{c^{2}}{2h\nu ^{3}}
\sum_{n=1}^{n=\infty} \frac{\mu(n)}{n} f(\frac{4\pi}{n}\kappa _{D}^{-1})\;,
\end{equation}
where $f$ is the inverse Laplace transform of the total photon power
spectrum.

\section{M\"{o}bius transform and supersymmetry}

In general, number theoretic methods promise to open up many possibilities
in various fields of mathematical physics \cite{J90}. Moreover, there is an
arithmetization trend of quantum physics.

We would like to mention that almost
simultaneously with Chen, Spector \cite{ds} pointed out
another application of the M\"{o}bius function. In his paper he showed
 the equivalence between
the M\"{o}bius function and the Witten topological index in
supersymmetric theories with discrete spectra. The Witten index
 operator $W=(-1)^{F}$
distinguishes fermionic from bosonic states and operators in
supersymmetric frameworks. It is just the number of fermionic
zero modes minus the bosonic zero modes (the spectral asymmetry).
In order to obtain the number theoretic
interpretation of the Witten index, Spector has made use of G\"{o}del
numbering in associating the prime numbers with the states of a
quantum system. Since $W$ is ill- defined, Witten \cite{W} suggested
to use a regularized generalized partition function
$\Delta=tr W e^{-\beta H}$ as a better order parameter for studying SUSY
breaking. Spector developed his arguments based only on
examples of systems with purely discrete spectra for which
 $\Delta$ is independent of $\beta$ as claimed by Witten.
However, it is known
that the continuous part of spectra causes the beta dependence of $\Delta$
\cite{ac}. Another
proof of Spector, again for systems with discrete spectra, concerns the
M\"{o}bius inverse transform, which shows up whenever one cancels some
of the bosonic degrees of freedom of a bosonic Hamiltonian by means of a
corresponding fermionic Hamiltonian, supersymmetrizing the bosonic
degrees of freedom to be deleted.


\newpage

\begin{thebibliography}{99}

\bibitem{03}
N.-x. Chen Phys. Rev. Lett. {64}, 1193 (1990); J. Math. Phys. {\bf 35},
3099 (1994);
N.-x. Chen, Y. Chen and G.-y. Li, Phys. Lett. A {\bf 149}, 357 (1990);
D. Xianxi, D. Jiqiong, preprints ITP-SB-90/104 and 105 (1990);
S.Y. Ren and J.D. Dow, Phys. Lett. A {\bf 154}, 215 (1991)
N.-x. Chen and G.B. Ren, {\em ibid} {\bf 160}, 319 (1991);
R.P. Millane, {\em ibid} {\bf A 162}, 213 (1992);
S.Y. Ren, {\em ibid} {\bf 164}, 1 (1992);
W. Deng and Y. Liu, {\em ibid} {\bf 168}, 378-382 (1992);
R.P. Millane, J. Math. Phys. {\bf 34}, 875 (1993);
A. Lakhtakia, Mod. Phys. Lett. B {\bf 5}, 491 (1991);


\bibitem{01}
J. Maddox, Nature (London) {\bf 344}, 377 (1990);
R.N. Gorgui-Naguib and  R.A. King in
{\em Mathematics in Signal Processing}, ed. T.S. Durrani et al.
(Clarendon Press, 1987);
I.S. Reed et al. , IEEE Trans. on Acoustics, Speech, and Signal Proc.
{\bf 38}, 458 (1990);
M.R. Schroeder, {\em Number Theory in Science and Communication
(with applications in cryptography, physics, digital information,
computing, and self-similarity)}, second enlarged edition,
(Springer, 1986). For a quite readable paper concerning the distribution
of prime numbers, see S. Guiasu, Math. Magazine {\bf 68}, 110 (1995)

\bibitem{02}
B.D. Hughes, N.E. Frankel and B.W. Ninham, Phys. Rev. A
{\bf 42}, 3643 (1990); B.W. Ninham et al.,
``M\"obius, Mellin, and mathematical physics",
Physica A {\bf 186}, 441-81 (1992)

\bibitem{Xie91}
T.L. Xie, F. Goldsmith, and W.M. Zhou, Astrophys. J.
{\bf 371}, L81 (1991); T.L. Xie et al., Astrophys. J. {\bf 402}, 216 (1993)

\bibitem{2}
S.W. Hawking, Nature (London) {\bf 248}, 30 (1974);
Commun. Math. Phys. {\bf 43}, 199 (1975)

\bibitem{GH82}
R. Geroch and J. Hartle, J. Math. Phys. {\bf 23}, 680
(1982); see also W. Israel, Lett. Nuovo Cim. {\bf 6}, 267 (1973)
\bibitem{Br91}
P.R. Brady, J. Louko and E. Poisson, Phys. Rev. D
{\bf 44}, 1891 (1991)

\bibitem{bch73}
J.M. Bardeen, B. Carter and S.W. Hawking,
Commun. Math. Phys. {\bf 31}, 161 (1973)

%\bibitem{th90} G. 't Hooft, Nucl. Phys. B {\bf 335}, 138 (1990)

\bibitem{b}
N.N. Bojarski, IEEE Trans. AP {\bf 30}, 778 (1982)

\bibitem{kj}
Y. Kim and D.L. Jaggard, IEEE Trans. AP {\bf 33}, 797 (1985)

\bibitem{p}
D.N. Page, Phys. Rev. D {\bf 13}, 198 (1976)

\bibitem{ma68}
R. Matzner, J. Math. Phys. {\bf 9}, 163 (1968)

\bibitem{s}
N. Sanchez, Phys. Rev. D {\bf 18}, 1030 (1978)

%\bibitem{mp} T.C. Mo and C.H. Papas, Phys. Rev. D {\bf 6}, 2071 (1972)

%\bibitem{f} R. Fabbri, Phys. Rev. D {\bf 12}, 933 (1975)

\bibitem{J90}
B. Julia, in {\em Number Theory and Physics}, Springer
Proc. in Physics {\bf 47}, 276 (1990), Eds. J.M. Luck, P. Moussa and
M. Waldschmidt.


\bibitem{ds}
D. Spector, Commun. Math. Phys. {\bf 127}, 239 (1990);
Phys. Lett. A {\bf 140}, 311 (1989). See also, M.J. Bowick, in
{\em Thermal Field Theories}, Eds. H. Ezawa, T. Arimitsu and Y. Hashimoto
(Elsevier, 1991), pp. 315-323

\bibitem{W}
E. Witten, Nucl. Phys. B {\bf 202}, 253 (1982)

\bibitem{ac}
R. Akhoury and A. Comtet, Nucl. Phys. B {\bf 246}, 253 (1984)

%\bibitem{rk} R. Kallosh, Phys. Lett. B {\bf 282}, 80 (1992)

%\bibitem{lav90} B.H. Lavenda, Intern. J. Theor. Phys. {\bf 29}, 1379
%(1990)

%R.R. Hall, J. reine angew. Math. {\bf 394}, 107-117 (1989)
%R.R. Hall, Mathematika {\bf 29}, 7-17 (1982)
%C. Hooley, Can. J. Math. {\bf XXV}, 1216-1223 (1973)
%R. Bellman and H.N. Shapiro, Duke Math. J. {\bf 21}, 629-37 (1954)


%%%%%%%%%%%%%%%%%%%%%%%%%%%%%%%%%%%%%%%%%%%%%%%%%%%%%%%%%%%%%%%%%%%%%%%%%

\end{thebibliography}
\end{document}